\documentstyle[prb,preprint,aps]{revtex}

\draft

\begin{document}
\title{Angular dependence of the bulk nucleation field $H_{c2}$ of aligned $MgB_{2}$
crystallites}
\author{O. F. de Lima, C. A. Cardoso, R. A. Ribeiro, M. A. Avila, and A. A. Coelho}
\address{Instituto de F\'{i}sica ''Gleb Wataghin'', UNICAMP, 13083-970, Campinas-SP,\\
Brazil}
\maketitle

\begin{abstract}
The angular dependence of the bulk nucleation field of a sample made of
aligned MgB$_{2}$ crystallites was obtained using dc magnetization and ac
susceptibility measurements. A good fitting of the data by the
three-dimensional anisotropic Ginzburg-Landau theory attests the bulk nature
of the critical field {\it H}$_{c2}$. We found a mass anisotropy ratio $%
\varepsilon ^{2}\approx 0.39$ that implies an anisotropy of the Fermi
velocity, with a ratio of 1.6 between the in-plane and perpendicular
directions, if an isotropic gap energy is assumed. For an s-wave anisotropic
gap this ratio could increase to 2.5. Besides the fundamental implications
of this result, it also implies the use of texturization techniques to
optimize the critical current in wires and other polycrystalline forms of MgB%
$_{2}$.
\end{abstract}

\pacs{74.25.Ha, 74.60.Ec, 74.60.Ge, 74.70.Ad}

Recent studies on the new MgB$_{2}$ superconductor \cite{cite1}, with a
critical temperature $T_{c}\ =39$ K, have evidenced its potential for
applications \cite{cite2,cite3}, although intense magnetic relaxation
effects limit the critical current density, $J_{c}$, at high magnetic fields 
\cite{cite4}. This means that effective pinning centers must be added \cite
{cite5} into the material microstructure, in order to halt dissipative flux
movements. Concerning the basic microscopic mechanism to explain the
superconductivity in MgB$_{2}$, several experimental \cite
{cite6,cite7,cite8,cite9,cite10,cite11,cite12} and theoretical \cite
{cite13,cite14,cite15} works have pointed to the relevance of a
phonon-mediated interaction, in the framework of the BCS theory. Questions
have been raised about the relevant phonon modes, and the gap and Fermi
surface anisotropies, in an effort to interpret spectroscopic and thermal
data that give values between 2.4 and 4.5 for the ratio $2\Delta _{0}/kT_{c}$%
, where $\Delta _{0}$ is the gap energy and $k$ is the Boltzmann constant.
Preliminary results on the $H_{c2}$ anisotropy have shown values of the
ratio between the in-plane and perpendicular directions which are around 1.7
for aligned MgB$_{2}$ crystallites \cite{cite16}, 1.8 for c-axis oriented
thin films \cite{cite17}, 2.6 for small single crystals \cite{Lee,M.Xu}, and
1.3 for very clean epitaxial thin films \cite{Alex}. \ Specific heat\cite
{cite10} and electron spin resonance\cite{simon} studies have also shown
broadening effects consistent with an $H_{c2}$ anisotropy. Here we present a
study on the angular dependence of $H_{c2}$ that points to a Fermi velocity
anisotropy around 2.5. Besides the fundamental aspects of this new result,
it also points clearly to the necessity of using texturization techniques to
optimize $J_{c}$ in wires and other polycrystalline components of MgB$_{2}$.

We measured a sample of well-aligned MgB$_{2}$ crystallites whose
preparation details have been described elsewhere \cite{cite16}. Briefly, a
MgB$_{2}$ powder of almost 100\% crystallites, having sizes up to $30\times
20\times 5$ $\mu $m$^{3}$, was obtained from a weakly sintered material
reacted at a temperature $T=1200$ $%
{{}^\circ}%
C$, much higher than the currently reported values below $900$ $%
{{}^\circ}%
C$. By spreading this powder on both sides of a paper we aligned the
crystallites with their $ab$ planes sitting on the paper surface. Several
samples were then mounted consisting of a pile of five squares of $3\times 3$
mm$^{2}$, cut from the {\it crystallite-painted} paper and glued with
Araldite resin.

Measurements of the magnetic moment and ac susceptibility were performed,
respectively, with a SQUID magnetometer (model MPMS-5) and a PPMS-9T
machine, both made by Quantum Design. In order to obtain the angular
dependence under an axial applied field a sample holder was built as
sketched in Fig. 1. All parts were machined from a teflon rod, except the
removable acrylic protractor, which is about 40 times larger than the hollow
box. The MgB$_{2}$ sample (in black) is mounted vertically inside the box
that is tightly inserted into a 5 mm diameter hole. This hole is drilled in
the plane surface of the sectioned rod that, finally, is attached at the end
of the system transport stick. A squared opening was made in the
protractor's center such that it fits precisely around the box sides. In
this way we are able to rotate the sample with a precision of $\pm $ 0.5
deg, which is good enough in view of the crystallites misalignment,
evaluated \cite{cite16} to be 2.3 deg around the sample $c$ axis. The
inconvenience of taking the sample holder out of the system, every time that
a new angle has to be set, is compensated by its simplicity and small
magnetic background.

Fig. 2 shows the magnetic field dependence of the magnetization in $T=25$ K,
for a few representative angles, $\theta $, between the sample $c$ axis and
the magnetic field direction. The inset displays a relatively sharp
transition with onset at $T_{c}=39$ K, measured with $H=10$ Oe in a
zero-field cooling (ZFC) and field-cooling measured on cooling (FCC)
procedures. The ZFC measurements shown in the main frame look noisy possibly
due to the effect of intense vortex creep \cite{cite4}, combined with a
complex regime of flux penetration in the granular sample. The occurrence of
random weak links and the varied coupling between grains produce a
fluctuating behavior in the sample overall response. However, in all cases
we were able to define $H_{c2}(\theta )$, at the crossing point between the
horizontal baseline and the straight line drawn across the experimental
points in the region near the onset of transition. This linear behavior of
the magnetization close to the onset is indeed expected from the
Ginzburg-Landau (G-L) theory \cite{cite18}. A constant paramagnetic
background was subtracted from all sets of data. In fact, one of the reasons
for measuring at 25 K is because at this temperature $H_{c2}(\theta )$
ranges between 28 - 36 kOe, where the paramagnetic background is saturated 
\cite{cite16}. $\ $Fig. 3 is a plot of ZFC followed by FCC magnetization
measurements (for $\theta $ = 85 deg and $T$ = 25 K) displaying clearly an
irreversibility field \ $H_{irr}\simeq 0.88\;H_{c2}$. This amount of
separation between $H_{irr}$\ and $H_{c2}$\ is similar to those observed in
c-axis oriented thin films\cite{cite17}, and contrasts with $H_{irr}\simeq
0.5\;H_{c2}$ observed for untextured bulk samples\cite{cite2,cite5,Finne}. \
Another significant difference is that the slope $dH_{c2}/dT$, close to $%
T_{c}$, is around 0.44 T/K for untextured bulk samples \cite{Finne}, while
it is between 0.2\ $\sim $ 0.3 T/K for textured samples \cite{cite16} and
single crystals\cite{M.Xu}. \ We believe that, besides the fact that MgB$_{2}
$ untextured samples give only an average response of its anisotropic
properties, the samples degree of purity might play also an important role,
since it affects the electronic mean free path.

Fig. 4 displays $H_{c2}(\theta )$ for $\theta $ between - 20 deg and 120
deg. The vertical error bars were estimated to be around $\pm $1 kOe while
the horizontal error bars, of $\pm $2.5 deg, almost coincide with the symbol
size. The solid line going through the experimental points represents a good
fit of the angular dependence, predicted by the 3D anisotropic G-L theory to
be \cite{cite18,cite19} $H_{c2}(\theta )=H_{c2}^{c}[\cos ^{2}(\theta
)+\varepsilon ^{2}\sin ^{2}(\theta )]^{-0.5}$, where $\varepsilon
^{2}=m_{ab}/m_{c}=(H_{c2}^{c}/H_{c2}^{ab})^{2}$ is the mass anisotropy ratio
and $H_{c2}^{c}/H_{c2}^{ab}$ is the ratio between the bulk nucleation field
along the $c$ direction and parallel to the $ab$ planes. We found $%
\varepsilon ^{2}\approx 0.39\pm 0.01$, giving $H_{c2}^{c}/H_{c2}^{ab}\approx
0.62$, which is close to the value of 0.59 anticipated \cite{cite16} by {\it %
ac} susceptibility measurements done for the two extreme $\theta $
positions, at 0 and 90 degrees. Fig. 4 shows also five data points ($\theta $
= 0, 25, 65, 85, 90 deg) marked with stars, which were obtained at the onset
of transition of the real part of the complex susceptibility, measured with
an excitation field of amplitude 1 Oe and frequency 5 kHz. From $%
H_{c2}^{ab}/H_{c2}^{c}=\xi _{ab}/\xi _{c}\approx 1.6$, $H_{c2}^{c}(T)=\Phi
_{0}/\left( 2\pi \xi _{ab}^{2}\right) $, and using the G-L mean field
expression \cite{cite18} for the coherence length $\xi (T)=\xi _{0}\left(
1-T/T_{c}\right) ^{-0.5}$, we find $\xi _{0ab}\approx 65$ \AA\ and $\xi
_{0c}\approx 40$ \AA , the coherence length at $T=$ 0 in the $ab$ planes and
along the $c$ axis, respectively. The quantum of flux, in CGS units, is $%
\Phi _{0}=2.07\times 10^{-7}$ G cm$^{2}$. The ratio $H_{c2}^{ab}/H_{c2}^{c}%
\approx 1.6$ reminds the relationship predicted for the surface nucleation
field \cite{cite20} $H_{c3}\approx 1.7$ $H_{c2}$. However, this is clearly
not the case of our data, as one can see from the expected angular
dependence of $H_{c3}(\theta )$ for thick samples \cite{cite21}, which is
plotted in Fig. 4 as a dash-dotted curve. The dashed curve in between
represents the well-known Tinkham's formula \cite{cite22} for the surface
nucleation field in very thin films. Therefore, a characteristic feature of
the surface nucleation field is the cusplike curve shape near $\theta $ = 90
deg, which contrasts with the sinusoidal shape followed by our data.

The macroscopic $H_{c2}$ anisotropy can be caused by an anisotropic gap
energy or by an anisotropic Fermi surface, as well as by a combination of
both effects \cite{cite23}. Assuming an isotropic gap, one gets $\xi
_{ab}/\xi _{c}=V_{F}^{ab}/V_{F}^{c}$, since\cite{cite18} $\xi \propto
V_{F}/\Delta _{0}.$\ Therefore, our data implies $V_{F}^{ab}\approx 1.6$ $%
V_{F}^{c}$, for the Fermi velocities within the $ab$ plane and along the $c$
direction. However, several experimental \cite{cite8,cite10,cite11} and
theoretical \cite{cite13,cite14,cite15} works have suggested an anisotropic
gap energy for MgB$_{2}$. In particular, two recent reports \cite
{cite11,cite15} rely on the analysis of spectroscopic and thermodynamic data
to propose an anisotropic s-wave pairing symmetry, such that a minimum gap
value, $\Delta _{0}\approx 1.2$ $kT_{c}$, occurs within the $ab$ plane.
Using this result and assuming an isotropic Fermi surface the expected $%
H_{c2}$ anisotropy would be \cite{cite15} $H_{c2}^{ab}/H_{c2}^{c}\approx 0.8$%
. This conflicts with our present findings and with other results\cite
{cite16,cite17} that show clearly $H_{c2}^{ab}>H_{c2}^{c}$. However, by
allowing a Fermi surface anisotropy in their model, Haas and Maki have found
that \cite{cite24} $V_{F}^{ab}\approx 2.5$ $V_{F}^{c}$ in order to match our
result of $H_{c2}^{ab}/H_{c2}^{c}\approx 1.6$, at $T=25$ K. Therefore the
two fundamental sources of microscopic anisotropy affect the $H_{c2}$
anisotropy of MgB$_{2}$ in opposite ways. As a consequence of combining both
effects to explain the $H_{c2}$ anisotropy, the Fermi velocity anisotropy
becomes about 60\% higher when compared with the isotropic gap hypothesis.
Interestingly, a calculation based on a two-band model has also found\cite
{cite25} $V_{F}^{ab}\approx 2.5$ $V_{F}^{c}$ , while a smaller value of $%
V_{F}^{ab}\approx 1.03$ $V_{F}^{c}$ was found in a band structure
calculation using a general potential method \cite{cite13}.

The relatively large scattering of reported values for the anisotropy ratio 
\cite{cite16,cite17,Lee,M.Xu,Alex} $H_{c2}^{ab}/H_{c2}^{c}$, varying between
1.3\ $\sim $ 2.6, could possibly be ascribed to at least three factors. The
first is the sample purity, since it affects directly the energy gap
anisotropy at the microscopic level \cite{cite19,cite23}. The second is the
experimental criterion used to define $H_{c2}$, since a reliable bulk
transition should be guaranteed. The third factor is that a possible
temperature dependent anisotropy ratio could arise from a temperature
dependent gap anisotropy \cite{cite15,cite24}. Therefore, results obtained
with samples of different purity levels and measured at different
temperatures most possibly should not produce the same anisotropy ratios.
This is clearly an area deserving much research work.

Concluding, the $H_{c2}$ anisotropy ratio for MgB$_{2}$ implies an
anisotropy ratio of the Fermi velocity of at least $V_{F}^{ab}/V_{F}^{c}%
\approx 1.6$, for an isotropic gap energy hypothesis. In a more realistic
scenario, of an s-wave anisotropic gap, this ratio could increase to $%
V_{F}^{ab}/V_{F}^{c}\approx 2.5$. \ Finally, since $J_{c}$ is proportional
to $\xi ^{2}$, it is worth noticing that \cite{cite26} $J_{c}$({\it H // c}) 
$/J_{c}$({\it H // ab}) $\approx \xi _{ab}/\xi _{c}\approx
H_{c2}^{ab}/H_{c2}^{c}$. Therefore, we anticipate that the in-plane critical
current density values are expected to be about 60\% higher than the values
along the c direction (H // ab). Indeed this $J_{c}$ anisotropy could be
even higher, as suggested by the larger $H_{c2}$ anisotropy observed in thin
films \cite{cite17} and single crystals \cite{Lee,M.Xu}. This means that, in
order to optimize $J_{c}$ in MgB$_{2}$ wires or other polycrystalline
components, some texturization technique will be required.

\bigskip

We thank S. Haas, A. V. Narlikar and O. P. Ferreira for useful discussions
and acknowledge the financial support from the Brazilian Science Agencies
FAPESP and CNPq.

\bigskip

\newpage

\begin{center}
FIGURE CAPTIONS
\end{center}

Figure 1 - Sketch of the sample holder used to get the angular dependence of
the sample magnetization in a SQUID magnetometer. The sample (in black) is
inside the hollow box, that is tightly inserted into a 5 mm diameter hole.
The removable acrylic protractor fits precisely around the box in order to
indicate the angular position.

\bigskip

Figure 2 - Zero Field Cooling magnetization measurements as a function of
the applied field, for $\theta $ = 0, 65, 80, 90 degrees. The bulk
nucleation field $H_{c2}(\theta )$ is defined at the crossing of the
auxiliary straight lines and the horizontal baseline $(M=0)$. The inset
shows ZFC and FCC magnetization measurements as a function of temperature
for $H=$ 10 Oe, giving $T_{c}$ $\approx $ 39 K.

\bigskip

Figure 3 - Zero Field Cooling (ZFC) followed by Field Cooling on Cooling
(FCC) magnetization measurements, for $\theta $ = 85 deg and $T$ = 25 K. \
The irreversibility field $H_{irr}$ and upper critical field $H_{c2}$ are
indicated by vertical arrows.

\bigskip

Figure 4 - Bulk nucleation field (or upper critical field), $H_{c2}$, as a
function of the angle, $\theta $ , between the sample $c$ axis and the
magnetic field direction. Plots of the expected angular dependence for the
surface nucleation field, $H_{c3}$, in thick samples (dash-dotted curve) and
very thin films (dashed curve) are also shown. The stars at $\theta $ = 0,
25, 65, 85, 90 degrees represent$H_{c2}(\theta )$ obtained at the onset of
transition of the real part of {\it ac} susceptibility measurements.

\label{Refs.}

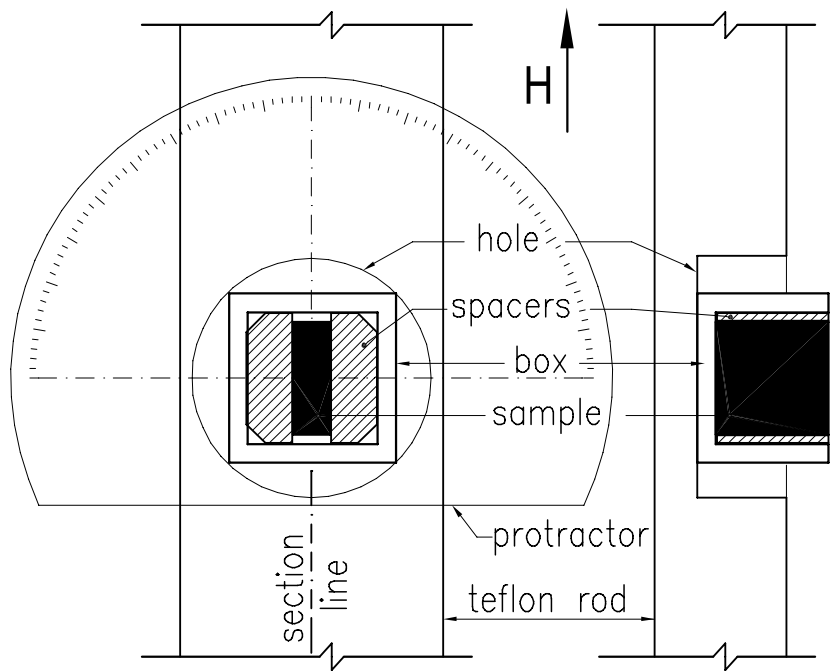
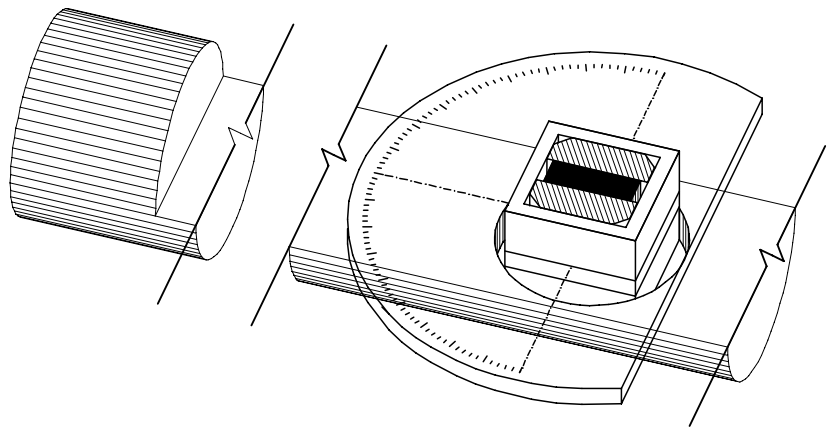
\end{document}